\documentclass{article}

\usepackage{amsmath}
\usepackage{color}
\usepackage{times}
\usepackage{graphicx}
\usepackage[numbers]{natbib}
\setcitestyle{maxnames=3}  % Limits author list in citations

\usepackage{lineno}
\usepackage{setspace}
\usepackage{gensymb}
\usepackage{siunitx}

\begin{document}

\title{Quantitative Analysis of Cell Membrane Tension in Time-Series Imaging and A Minimal Lattice Model of Single Cell Motion}
\author{Hiroki Nishitani$^1$, Takashi Miura$^{2}$}
%\address{$^1$九州大学医学部, $^2$九州大学大学院医学研究院}

\maketitle

%\begin{multicols}{3}
\doublespacing

%\linenumbers
\section{Abstract}

% 数理生物学の分野で、細胞形状を再現するモデルとして体積保存項付きPhase field法がよく用いられる\cite{Nonomura2012, Imoto2021}。このモデルでは表面張力が導入されているが、実際の細胞でも細胞膜やその直下のcortical actinによって表面張力が生じることが知られている\cite{Cartagena-Rivera2017}。本研究では、時系列データから表面張力の推定を行う手法を考案した。細胞質及び核を蛍光ラベルしたMDCK細胞を用いてタイムラプス画像を取得し、画像処理で細胞境界を二値化した画像を作成した。次に、時刻tの境界格子の法線方向を計算し、法線方向に動くと仮定して時刻t+dtで対応する境界格子を指定した。この格子のペアを用いて界面の法線方向の速度$V$を算出した。さらに、円板状のカーネル関数を用いた畳み込みを用いて曲率 $\kappa$を算出し、曲率 - 速度の散布図を線形回帰した。
% Phase-field方程式から界面の発展方程式V=σκを求めることができる。ここでσは表面張力に粘性係数を掛けたものだが←消す
% これを実効表面張力として$\sigma$の推定を試みた。実測値では曲率–速度関係は線形にはならず、折れ線やS字曲線となった。この原因を探るため、単細胞の運動を記述するミニマルな格子モデルを作成し、曲率–速度関係で同様のパターンが生じることを見いだした。

Cell membrane tension directly influences various cellular functions. In this study, we developed a method to estimate surface tension from time-series data. We obtained the curvature-velocity relationship from time-series of binarized cell shape images, and the effective surface tension term was calculated from linear regression. 

During the process, we observed an S-shaped pattern in the curvature-velocity relationship. To understand the dynamics, we constructed a minimal lattice model describing single-cell motion. The model consists of surface tension and protrusion formation, and the characteristic parameters are obtained from experimental observations. We found that similar patterns emerged in the curvature-velocity relationship.

%From the phase-field equation, the interface evolution equation can be expressed as \( V = -\sigma \kappa \). 
%Here, \( \sigma \) represents the product of surface tension and viscosity. 
%In this study, we treated \( \sigma \) as an effective surface tension and attempted to estimate its value. However, in the experimental data, the relationship between curvature and velocity was not strictly linear but exhibited piecewise linear or S-shaped patterns. To investigate the underlying mechanism, we constructed a minimal lattice model describing single-cell motion and found that similar patterns emerged in the curvature-velocity relationship.

\section{Introduction}
% 参考
% https://www.sci.kyushu-u.ac.jp/koho/qrinews/qrinews_210609.html

% https://seikagaku.jbsoc.or.jp/10.14952/SEIKAGAKU.2017.890508/data/index.html?utm_source=chatgpt.com

% https://www.cell.com/trends/cell-biology/fulltext/S0962-8924(12)00115-8?_returnURL=https%3A%2F%2Flinkinghub.elsevier.com%2Fretrieve%2Fpii%2FS0962892412001158%3Fshowall%3Dtrue

\subsection{Shape of cultured cells}
% 細胞形状の動的な変化は、細胞の運動能力に深く関わっており、発生、免疫応答、がんの転移といった過程に不可欠である。
% 細胞はさまざまな種類の突出構造を持ち、平坦なラメリポディア（lamellipodia）、指状のフィロポディア（filopodia）、丸いブレブ（blebs）といった構造がある。
% フィロポディアは、細胞移動、微小管の誘導、シグナル伝達に関与する細長い繊維状の構造である \cite{Gallop2020}。
% ラメリポディアとフィロポディアはアクチン骨格の重合によって膜が押されて出来る構造であり、アクチン骨格はリンカータンパク質を介して細胞膜と繋がっている。
% ラメリポディアは薄く広がっていて基質との接着性が高く、2次元の基質上の運動において主要な突出構造として広く研究されてきた。
% ブレブは細胞骨格と細胞膜の結合が壊れた部位で、静水圧により膜を押し出して生じる構造である。
% ブレブは基質との接着性が弱く、3次元の環境下で柔軟かつ高速な運動を可能にするとして近年注目されている。
% 突出構造は運動性だけでなく、基質を探索するセンサーとしても機能すると考えられており、細胞形状の果たす役割は大きい。

% 膜の張力の相対的な大きさ?でぐねぐねとか扇形？になるやつ、でもfigure2があんま説明なくていまいち理解が及んでない
% MDCK細胞は上皮細胞のモデルとして研究されており、集団でシート状の構造を形成することなどが知られている。 書く？MDCKのレビュー論文とか見つからない

Dynamic changes in cell shape are closely related to cellular motility and are essential for processes such as development, immune responses, and cancer metastasis \cite{Bodor2020}. Cells possess a variety of protrusive structures, including flat lamellipodia, finger-like filopodia, and round blebs. Lamellipodia are thin, wide structures with strong adhesion to the substrate and have been widely studied as the primary protrusive structures involved in two-dimensional substrate-based motility \cite{Innocenti2018}. Filopodia are long, fiber-like structures that work during cell migration, microtubule guidance, and signal transduction \cite{Gallop2020}. Lamellipodia and filopodia are formed by the polymerization of the actin cytoskeleton, which pushes the membrane outward. Blebs, on the other hand, arise from regions where the connection between the cytoskeleton and the membrane is disrupted, allowing hydrostatic pressure to push the membrane outward. Unlike lamellipodia, blebs exhibit weak adhesion to the substrate and have recently gained attention for their role in enabling flexible and rapid movement in three-dimensional environments. Protrusive structures not only contribute to cellular motility but are also thought to function as sensors for exploring the substrate. Thus, the role of cell shape in cellular function is substantial.

%Madin-Darby Canine Kidney (MDCK) cells are widely used as a model for epithelial cells and are known to form sheet-like structures as a collective.

\subsection{Theoretical models of cell shape}

There are several theoretical models that can reproduce single-cell shapes.

\subsubsection{Phase field model}

The phase-field model is a continuous field model that describes a variable \( \phi(\mathbf{r}, t) \) representing phases \cite{Kobayashi:1993ww}. Using this method, we can express the movement of the interface with surface tension. This model has been utilized to express single and multicellular dynamics \cite{Nonomura2012, Imoto2021}. To describe cells, they introduce active cell movement (cell kinesis), chemotaxis, cell adhesion, and volume conservation to the model by defining appropriate energy.

\subsubsection{Cellular Potts model}
% 離散的に細胞運動を再現するモデルとしては、cellular Potts modelが知られている。これは、細胞の接着力や体積保存、化学走化性などからなるエネルギーを定義し、マルコフ連鎖モンテカルロ法を用いて時間発展させるモデルである。CPMは単純で計算が簡単であり、拡張性も高いことから広く使われている。しかし、パラメータと細胞の物理的な定数を対応させることや、モンテカルロ法を時間ダイナミクスの表現とみなすのが難しいといった問題がある。

A well-known discrete model for reproducing cell motion is the cellular Potts model (CPM) \cite{Graner1992}. This model represents cell shape as a pattern on a lattice, and defines an energy functional
\[
E = E_{\text{adhesion}} + E_{\text{volume}} + E_{\text{chemotaxis}}
\].
$E_{\text{adhesion}}$, $E_{\text{volume}}$, and $E_{\text{chemotaxis}}$ represent cell surface tension, volume conservation, and polarized cell movement, respectively. Then the cell shape evolves over time according to the functional using the Markov chain Monte Carlo method. The CPM is simple, computationally efficient, and highly extensible, making it widely used. However, it faces challenges such as mapping model parameters to physical constants of cells and interpreting the Monte Carlo method as representing temporal dynamics.

\subsection{Surface tension of cell cortex}
% 細胞膜の張力は、膜輸送といった細胞機能や細胞の形状、運動性に直接関わっていることが明らかにされており、重要な概念である。
% 膜張力は、細胞膜から引き出された膜テザーを維持するために必要な力として測定することができる。
% また細胞に関する張力は複数あり、membrane tension、cortical  tension、 tension on  the substrateなどがある。
% membrane  tensionは、脂質二重層内の静水圧による張力と、膜と細胞骨格の接着の大きく2つの要素から構成され、これら2つは互いに影響している。
% cortical  tensionは細胞膜にくっついたアクチン細胞骨格が、緩んだり収縮したりすることで生じ、membrane tensionに大きく影響する。
% tension on  the substrateは、アクチン細胞骨格が基質に強く結合することで生じ、membrane tensionへの影響は小さい。
% こうした張力の大きさは、細胞の様々な形状(突出・退縮部位を交互に境界にもつ細胞や、前縁にラメリポディアを形成して後縁が収縮する極性をもつ細胞など)を特徴づけている。

Cell membrane tension, that are involved in both models, is a critical concept that has been shown to directly influence cellular functions such as membrane trafficking, as well as cell shape and motility \cite{Gauthier2012}. Membrane tension can be measured as the force required to maintain a membrane tether pulled from the cell surface. Direct measurement of membrane tension has been done using atomic force microscopy and optical tweezers \cite{Cartagena-Rivera2017,Bambardekar2015}

There are multiple types of tension associated with cells, including membrane tension, cortical tension, and tension on the substrate \cite{Gauthier2012}. Membrane tension consists of two major components: hydrostatic tension within the lipid bilayer and adhesion between the membrane and the cytoskeleton, both of which influence each other. Cortical tension arises from the relaxation or contraction of the actin cytoskeleton attached to the cell membrane and has a significant impact on membrane tension. In contrast, tension on the substrate is generated by the strong adhesion of the actin cytoskeleton to the substrate, but its influence on membrane tension is minimal. 
The magnitudes of these tensions characterize various cell shapes, such as cells with alternating protrusive and retractile regions at their boundaries or polarized cells with lamellipodia formation at the leading edge and contraction at the trailing edge.

\subsection{Research summary}
% 我々は、タイムラプス画像の解析と数理モデルを用いて、MDCK細胞の単細胞での細胞膜の変形について議論する。
% 我々は、細胞のタイムラプス画像を二値化して、界面の突出や退縮に関する時空間的な特徴を調べた。また、界面の曲率と法線方向の速度の関係から界面張力について調べた。
% 我々は、界面の変形に関する物理的なパラメータを直接モデルに組み込み、細胞の形状を再現する離散モデルを考案した。

We examined the deformation of the cell membrane in single MDCK cells using time-lapse image analysis and mathematical modeling. By image processing, we analyzed the spatiotemporal characteristics of interface protrusion and retraction. Using this data, we investigated surface tension by examining the relationship between the curvature of the interface and the velocity in the normal direction. Finally, we developed a discrete model that directly incorporates physical parameters related to interface deformation, allowing us to accurately reproduce the shape of cells.

\section{Materials and methods}
\subsection{Cell culture}
Madin-Darby Canine Kidney (MDCK), which constitutively expresses EGFP, was kindly gifted from Professor Hideru Togashi (Kobe University). Cells were maintained in Dulbecco's Modified Eagle Medium (DMEM) + 10 \% fetal bovine serum (FBS) + 1 \% antibiotics (Gibco). Cells were cultured on a plastic dish and passaged once per week.

\subsection{Time-lapse observation}
Subconfluent MDCK cells are detached from the culture dish using 0.05 \% Trypsin-EDTA. Cells are then seeded on a 35 mm glass-bottom dish (Matsunami glass). The cells were allowed to attach to a glass surface for two hours in a CO2 incubator. Then, the dish was set in a stagetop CO2 incubator (Tokai Hit), and the cell dynamics were observed using a Nikon A1R confocal microscope (Nikon). Cell images were taken every 1 minute for 3 hours. 

In some cases, we added 2 $\mu M$ cytochalasin D and/or 2.5 $\mu M$ colchicine to the culture medium to observe the effect of cytoskeleton disruption.

\subsection{Image processing}
\subsubsection{Binarization}
Captured nd2 files are converted to NumPy array using nd2reader, and processed using Python. Green channels (cytoplasm) were separated, and smoothened using OpenCV \cite{itseez2015opencv}, and obtained mask image by binarizing the smoothened image. A manually defined threshold was used. 

\subsubsection{Measurement of curvature by kernel}
% マスク画像に対して、Scipyのconvolve2d関数を用いて円盤状のカーネルと畳み込み処理を行った \cite{Virtanen2020}。次に、カーネルと半径 1/κ の円との交差領域の面積を計算することで、畳み込み結果を曲率 κ に変換する関数を作成した。カーネルより小さすぎる円は対象外とした。
% マスク画像を収縮（エロージョン）および膨張（ダイレーション）処理することで、収縮画像（eroded image）および膨張画像（dilated image）を生成した。内側境界（inner boundary）の画像は、マスク画像から収縮画像を差し引くことで得られ、外側境界（outer boundary）の画像は、膨張画像からマスク画像を差し引くことで得られた。
% 界面ピクセルの曲率は、畳み込まれたマスク画像と内側境界画像を乗算することで求められ、最終的に畳み込み結果を曲率 κ に変換した。

\begin{figure}[h]
    \centering
    \includegraphics[width=0.6\linewidth]{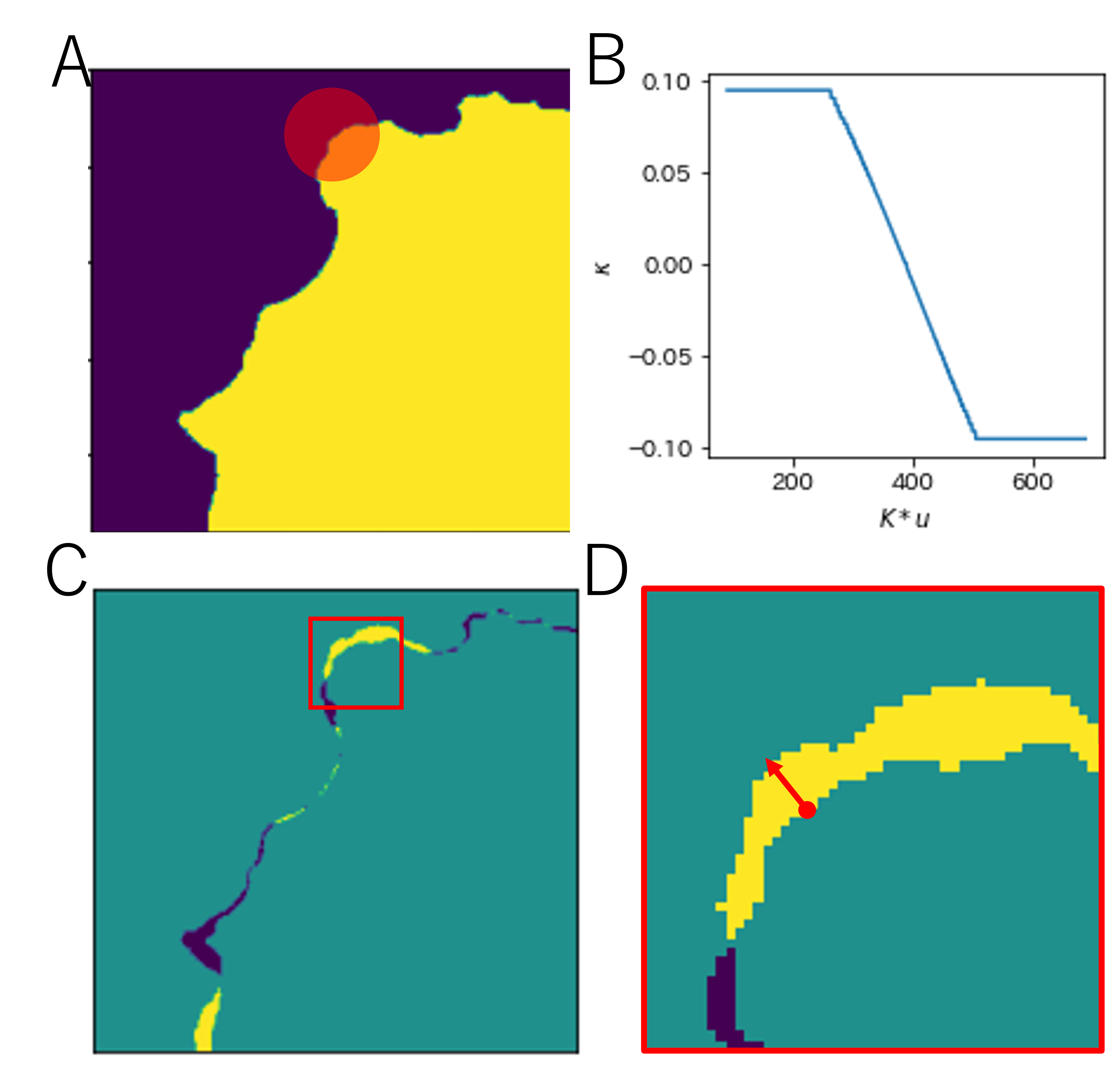}
    \caption{A conceptual diagram of curvature measurement. (A) Measurement of curvature using a circular kernel. The area of the intersection of kernel $K$ (red) and cell $u(x,y,t)$ (yellow) was measured. (B) Relationship between convolution area $K*u$ and curvature $\kappa$. There is a correlation between the area $K*u$ of overlap between the kernel and the cell and the curvature $\kappa$. (C) Difference $u(x,y,t+1)-u(x,y,t)$ of the binary cell image between two frames. (D) An enlarged view of (c). Red arrow indicates $V \bf{n}_{i,t}$}
    \label{fig:kernel_curvature}
\end{figure}

A convolution operation was performed on the mask image using a disk-shaped kernel with the \texttt{convolve2d} function in SciPy \cite{Virtanen2020} (Fig. \ref{fig:kernel_curvature}a). Subsequently, a function was developed to convert the convolution result into curvature \( \kappa \) by calculating the area of intersection between the kernel and a circle with a radius of \( 1/\kappa \) (Fig. \ref{fig:kernel_curvature}b). Circles significantly smaller than the kernel were excluded from the analysis.

%Erosion and dilation operations were applied to the mask image to generate the eroded image and the dilated image, respectively. The inner boundary image was obtained by subtracting the eroded image from the mask image, while the outer boundary image was obtained by subtracting the mask image from the dilated image.  
Erosion and dilation were applied to the mask image. %generating eroded and dilated images. 
The inner boundary was obtained by subtracting the eroded image from the mask, and the outer boundary by subtracting the mask from the dilated image.
The curvature of the interface pixels was computed by multiplying the convolved mask image with the inner boundary image, and the final curvature \( \kappa \) was determined from the convolution result.

\begin{figure}
    %\centering
    \includegraphics[width=\linewidth]{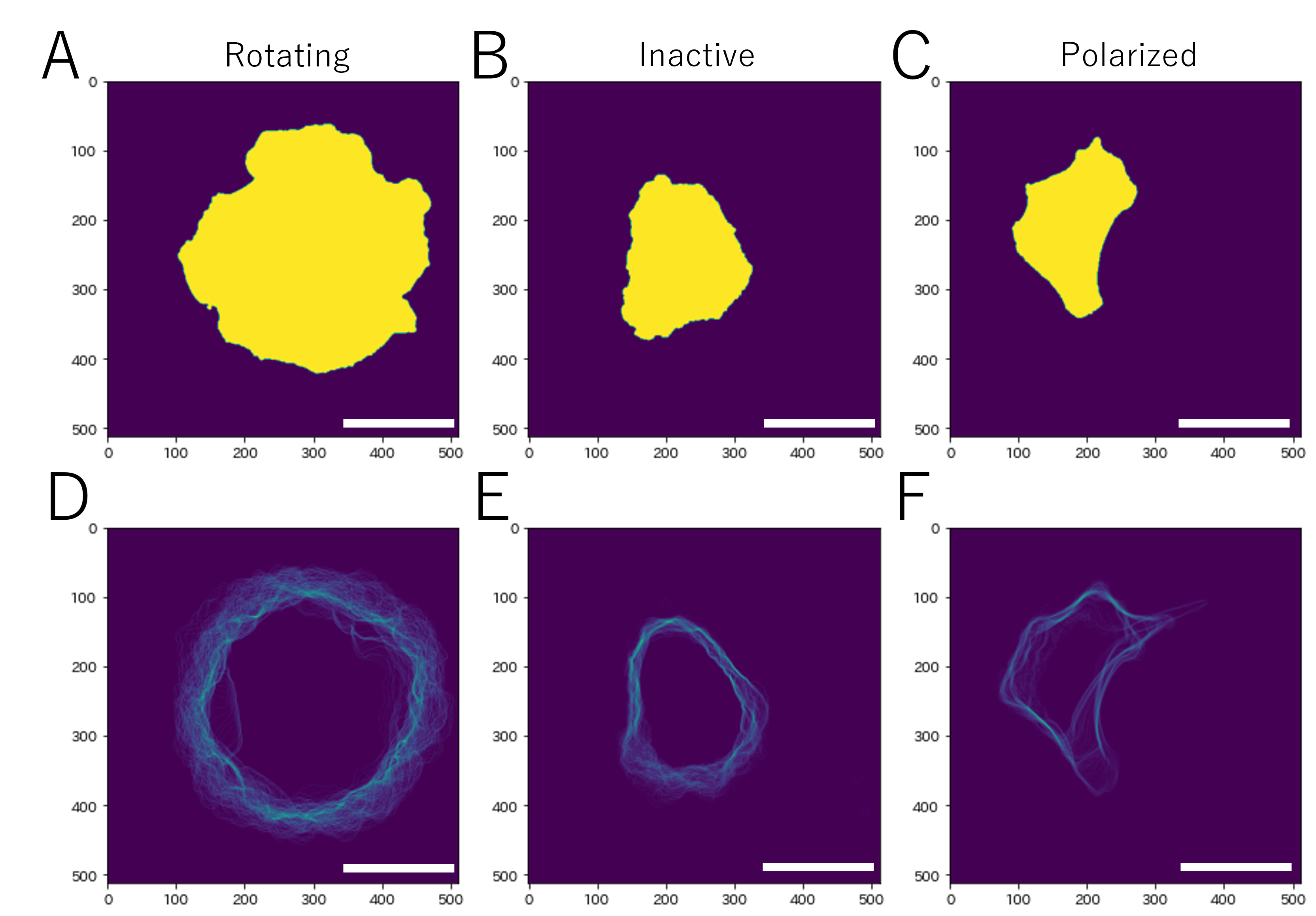}
    \caption{Three classes of cell movement. (A) A cell with alternating protrusive and retractile regions at the cell boundary. (B) A cell with a relatively inactive boundary. (C) A cell with lamellipodia formation at the leading edge and contraction at the trailing edge. (D-F) Time-series overlay of the cell boundaries from (A-C). Scale bars = \SI{500}{\micro \metre}.}
    \label{fig:cell_example}
\end{figure}

\subsubsection{Quantification of interface velocity}
% 界面速度は以下の手順で算出した。まず、時刻 \( t \) における内側境界の座標リストを \( \mathbf{r}_{i,t} \) として取得した。次に、ndimage の \texttt{gaussian\_filter} 関数を用いて平滑化画像 \( s(x,y) \) を得た。\( i \) 番目の内側境界点における外向き単位法線ベクトル \( \mathbf{n}_{i,t} \) は、  
% \[
% \mathbf{n}_{i,t} = \frac{\nabla s(x,y)}{|\nabla s(x,y)|}
% \]  
% として求めた。次に、時刻 \( t+dt \) における内側境界の座標を取得した。最後に、\( j \) を選択し、以下を最小化することで、\( \mathbf{r}_{i,t} \) に対応する座標 \( \mathbf{r}_{j,t+dt} \) を決定した。  
% \[
% |\mathbf{r}_{i,t} + V \mathbf{n}_{i,t}-\mathbf{r}_{j,t+dt}|
% \]

Interface velocity $V$ (Fig. \ref{fig:kernel_curvature}c, d) was calculated as follows: at first, we obtained a list of coordinates of the inner boundary at a certain timepoint $t$ as $\mathbf{r}_{i,t}$. Next, we obtained a smoothened image $s(x,y)$ using a gaussian\_filter function in ndimage. Outward unit normal vectors at $i$th inner boundary points $\mathbf{n}_{i,t}$ were obtained by $\nabla s(x,y)/|\nabla s(x,y)|$. Next, we obtained inner boundary coordinates at time $t+dt$. Finally, we chose a coordinate $\mathbf{r}_{j,t+dt}$ that corresponds to $\mathbf{r}_{i,t}$ by choosing $j$ that minimizes $|\mathbf{r}_{i,t} + V \mathbf{n}_{i,t}-\mathbf{r}_{j,t+dt}|$.

\subsubsection{Quantification of protrusion dynamics}
% inner boundary at certain timepoint $t$に対して、重心からの角度と長さを計算した
% 角度は整数値に丸めて、同じ角度をもつ格子を平均することで、0〜360°の角度に対して重心からの界面の長さを得た
% 時間方向に差分を取り、距離1以上長さが増えたもの、距離1以上長さが減ったもの、距離が1以下しか変化しなかったもの、で1、-1、0に3値化した
% これにより、重心からの角度vs時間の二次元配列について、突出、退縮、停滞を割り当てた
% この二次元配列について時空間の特徴長さなどを調べることで、protrusion dynamicsを解析した
% それぞれの角度について、時間方向に1が連続する数をカウントすることで、protrusionの持続時間のヒストグラムを得る。
% また、それぞれの時間について、空間方向に1が連続する数をカウントすることで、突出の幅のヒストグラムを得る。

The angle and distance from the centroid to the inner boundary at a certain timepoint were calculated.
The angles were rounded to integer values, and the grid points with the same angle were averaged to obtain the interface length from the centroid for angles ranging from 0\degree  to  360\degree.
The temporal differences in this length were calculated, and based on these differences, the changes were categorized into three states:
\begin{enumerate}
 \item Values were assigned as 1 for increases in length of 1 or more units.
 \item Values were assigned as -1 for decreases in length of 1 or more units.
 \item Values were assigned as 0 for changes within 1 unit.
\end{enumerate}
Using this method, a two-dimensional array representing angle vs. time was generated, where the states of protrusion, retraction, and stasis were assigned to each point.
For each angle, the number of consecutive 1s in the time direction was counted to obtain the histogram of protrusion duration.
Additionally, for each timepoint, the number of consecutive 1s in the spatial direction was counted to obtain the histogram of protrusion width.

%\newcolumn

\subsubsection{Measurement of surface tension}
% 界面張力の計測
% アクチン骨格が細胞膜を押すことで突出が生じるため、突出の影響を除くことで表面張力を推定することができる。
% 極性が強くない、突出・退縮が回転伝播している細胞を選んだ。
% 突出部分の判定には、上の解析で作成した、突出退縮を表す時空間の配列を用いる。
% 曲率vs速度のplotを作成する過程で、重心からの角度を計算し、退縮部分のみを用いた。

Since protrusions are generated by the actin cytoskeleton pushing against the cell membrane, surface tension can be estimated by excluding the influence of protrusions. To achieve this, we selected cells in which protrusions and retractions propagated rotationally.  

The identification of protrusive regions was based on the spatiotemporal array representing protrusion and retraction, constructed in the preceding analysis. During the process of generating the curvature vs. velocity plot, angles from the centroid were computed, and only the retraction regions were utilized.

\subsection{Numerical simulation}

We implement both the lattice-based model and the phase field model using Python and Julia. Details of the model are described in the Supporting Information.

\section{Results}
\subsection{Quantitative measurement of cell shape dynamics}

\subsubsection{Observation of cell movement}

We obtained time-lapse movies of cells using Nikon A1R confocal microscope. Movements of a total of 3 hours were recorded for each cell. In the experiment, the following types of cells were observed:  
\begin{itemize}
\item A cell with alternating protrusive and retractile regions at its boundaries (Fig. \ref{fig:cell_example}A, D).  
In this cell, the protrusions were prominent.  
\item A cell with relatively inactive boundaries (Fig. \ref{fig:cell_example}B, E).  
In this cell, the protrusions were moderately large, but no rotational propagation was observed.  
\item A cell with lamellipodia formation at the leading edge and contraction at the trailing edge (Fig. \ref{fig:cell_example}C, F).  
In this cell, protrusions were localized at the leading edge and appeared fine in structure. 
\end{itemize}

% 実験では以下のような細胞があった。
% A cell with alternating protrusive and retractile regions at their boundaries. 
% この細胞では、突出が大きかった。
% A cell with a relatively inactive boundaries. 
% この細胞では、突出がやや大きく、回転はしていなかった。
% A cell with lamellipodia formation at the leading edge and contraction at the trailing edge. 
% この細胞では突出はleading edgeに限られ、細かかった。
% (D-F) Time-series overlay of the cell boundaries from (A-C).

\subsubsection{Quantification of interface movement}
%細胞膜の界面の運動性を、2frame間の細胞画像の差分面積を周長で割ったものの時間平均で評価する。
% 細胞骨格阻害剤が界面の運動性に及ぼす影響を評価するため、control、アクチン重合阻害剤であるcytochalasin D処理群、微小管重合阻害剤であるcolchicine処理群、両阻害剤を併用した群（cytochalasin D + colchicine）の4群について、得られた界面の運動性の分布をボックスプロットで可視化した（図A）。

% 図Aに示すように、4群間で界面の運動性にばらつきが見られたため、まずKruskal-Wallis検定を行ったところ、統計量H=33.9657、有意確率p=2.0145×10^{−7}となり、有意水準5%のもとで4群の中央値に有意差があると判断された。そこで、各ペア間の差をTukey法で解析した。その結果、control群とcytochalasin D群、control群とcytochalasin D + colchicine群、colchicine群とcytochalasin D群、colchicine群とcytochalasin D + colchicine群のペア間で有意差が認められた。一方で、control群とcolchicine群の間、およびcytochalasin D群とcytochalasin D + colchicine群の間では統計的に有意な差は認められなかった。これらの結果から、アクチン重合阻害剤であるcytochalasin Dを処理した群では、コルヒチン群やコントロール群と比べて界面の運動性が顕著に低下し、コルヒチンとの併用群でも同様の傾向が示唆された。一方、コルヒチン単独ではコントロール群と大きな差が見られないことから、アクチン線維の動態が本指標に対してより大きな影響を及ぼす可能性が考えられる。

\begin{figure}
    \centering
    \includegraphics[width=0.6\linewidth]{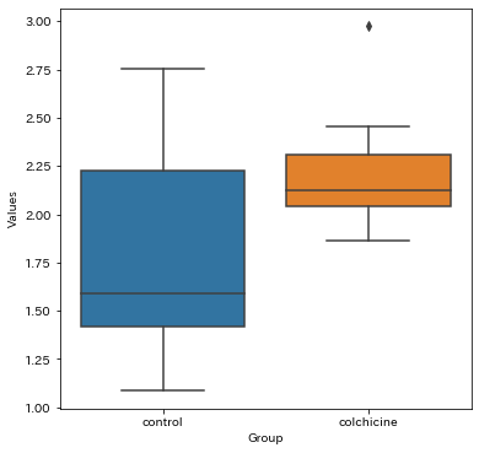}
    \caption{Box plot of interface motility between the two experimental groups.}\label{fig:paper_border_index_drug}
\end{figure}

Motility of the cell membrane interface was quantified as the time-averaged ratio of the difference in area between two consecutive frames to the perimeter. To assess the effect of cytoskeletal inhibitors on interface motility, we compared the control group and the colchicine-treated group (a microtubule polymerization inhibitor)  (Fig. \ref{fig:paper_border_index_drug}). We could not detect a statistically significant difference. 

%These results suggest that cells treated with cytochalasin D exhibited significantly reduced interface motility compared to both the colchicine-treated and control groups, and a similar trend was observed in the cytochalasin D + colchicine-treated group. On the other hand, colchicine alone did not induce a significant change compared to the control group. This finding implies that actin filament dynamics play a more substantial role in determining this motility index than microtubule dynamics.

\vspace{15mm}

\subsubsection{Curvature-velocity relationship.}
% 実験において計測した、曲率vs界面の速度のplotを示す。複数細胞ではplotは折れ線に近く、凸部分で傾きが小さくなった。単細胞では、S字のようなグループ、中間的なグループ、折れ線に近いグループに別れた。
% また、各グループで界面の運動性を比較すると、曲率vs界面速度関係がs字のグループで運動性が大きかった。
% S字グループは突出が盛んで回転伝播している細胞であり、表面張力に対する突出が相対的に強く、一定の大きさをもつことがplotの形状を決めていると考えられる。
% 中間的なグループと折れ線のグループのplot形状の違いついては、界面の運動性や突出の大きさが異なることが原因として考えられる。
% 多数の細胞を平均すると、大きさや頻度の異なる突出により、特定の曲率への影響が薄まり、曲率0付近で折れた折れ線のような形になると考えられる。

\begin{figure}
    %\centering
    \includegraphics[width=\linewidth]{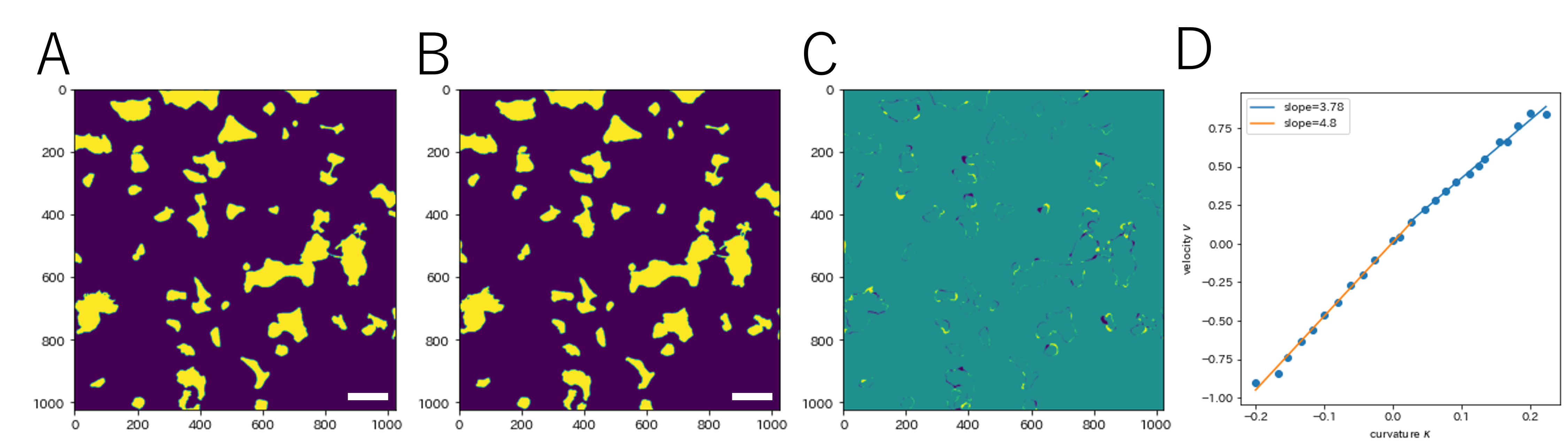}
    \caption{Observation of curvature-velocity relationship of multiple cells. 
    %(A, B) Patterns of multiple cells observed in the experiment. \( u(t) \): the binarized cell image at time \( t \), where the cell region is represented as 1 and the non-cell region as 0. 
    (A) The cells at \( t = 0 \).  (B) The cells at \( t = 1(300 sec) \).  (C) The difference image between \( t = 1 \) and \( t = 0 \).  (D) The relationship between curvature \( \kappa \) and velocity \( V \) obtained from multiple cell images in the experiment. The plot forms a piecewise linear shape, with a change in slope near \( \kappa = 0 \).  
Scale bar: \( 200 \:\mu\text{m} \).
}
    \label{fig:experiment_multicell}
\end{figure}

\begin{figure}
    %\centering
    \includegraphics[width=\linewidth]{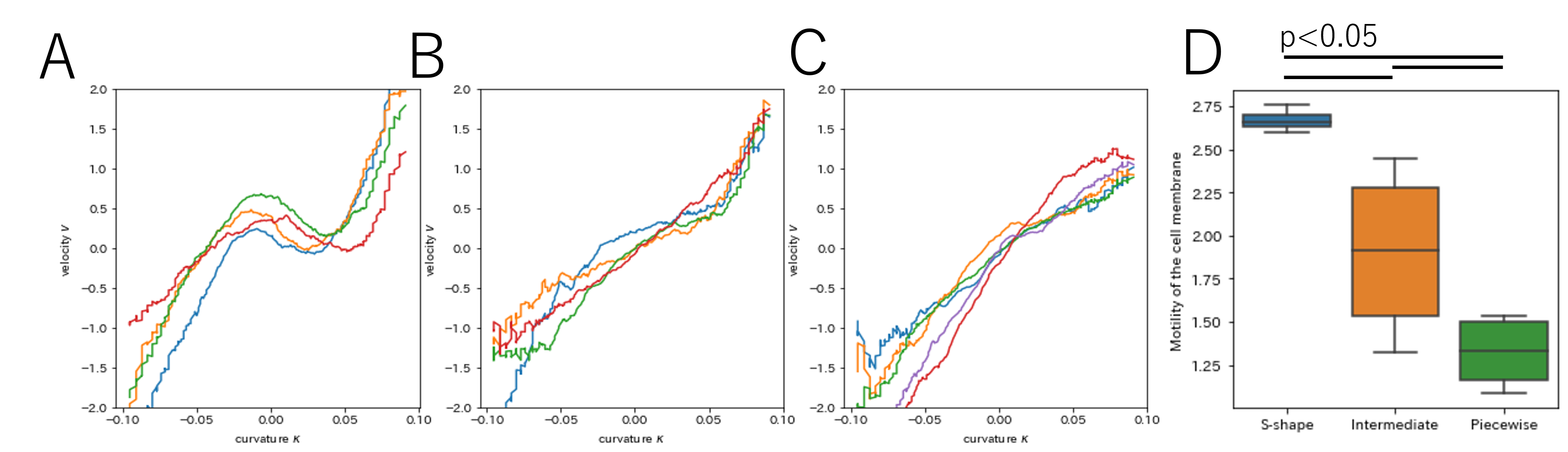}
    \caption{Curvature vs. interface velocity measured in single cells. (A) S-shape pattern. (B) Intermediate pattern. (C) Piecewise linear pattern. (D) Membrane interface Motility of (A-C).}\label{fig:paper_singlecell_curvature_vs_velocity}
\end{figure}

We obtained the curvature vs. interface velocity plot from experimental measurements. In the case of multiple cells, the plot exhibited a piecewise linear pattern, with a reduced slope in the convex regions (Fig. \ref{fig:experiment_multicell}). For single cells, three distinct groups were identified: an S-shaped group, an intermediate group, and a piecewise linear group (Fig. \ref{fig:paper_singlecell_curvature_vs_velocity}A-C).  

A comparison of interface motility across these groups revealed that the S-shaped group exhibited the highest motility (Fig. \ref{fig:paper_singlecell_curvature_vs_velocity}D). The S-shaped group corresponds to cells with active protrusions and rotational propagation. In these cells, protrusions are relatively strong compared to surface tension and maintain a characteristic size, which is likely to determine the shape of the plot.  

The differences in plot shapes between the intermediate and piecewise linear groups may be attributed to variations in interface motility and protrusion size. When averaging across multiple cells, the influence of protrusions with different sizes and frequencies is diminished, resulting in a piecewise linear shape with a kink near zero curvature.

\vspace{1cm}
\vspace{15mm}

% \begin{figure}
%     %\centering
%     \includegraphics[width=\linewidth]{_Figures/hirosima_experiment_onecell.pdf}
%     \caption{(A,B)実験で得られた単細胞のタイムラプス画像。$u(t)$: 時刻$t$における2値化細胞画像。細胞部分で1, 非細胞部分で0。(A)$t=0$での細胞画像。(B) $t=1$での細胞画像。(C) $t=1$と$t=0$の差分画像。(D,E)実験データから計測した$\kappa$ vs $V$関係。(D) $\kappa$ vs $V$がS字カーブになるもの。(E) $\kappa$ vs $V$が折れ線になるもの。(F)A,Bの例での$V$の絶対値の平均。スケールバー： $50\:\mu m$。}
%     \label{hirosima_experiment_onecell}
% \end{figure}

\subsubsection{Quantification of protrusion formation dynamics}
% a two-dimensional array representing angle vs. time was generated, where the states of protrusion, retraction, and stasis were assigned to each pointを示す。突出は黄色、退縮は黒、停滞は緑で表される。
% Rotating cellでは、突出部分について、時空間的に一定の特徴長さをもち、突出は一定の速度で界面を回転して伝播している。histogram of protrusion durationとhistogram of protrusion widthを示す。classで重み付けしたfrequencyを見ると、時間の特徴長さとして5frame、空間の特徴長さとして20°ほどにピークがある。これらは突出の持続時間と幅として解釈できる。
% Inactive cellでは、突出の特徴長さと個数、回転が少し弱まっている。
% Polarized cellでは尾部の退縮が180°ほどに表れており、他の部分の突出は細かい。
% colchicine処理群では、突出は存在するものの細かかった。

\begin{figure}
    %\centering
    \includegraphics[width=\linewidth]{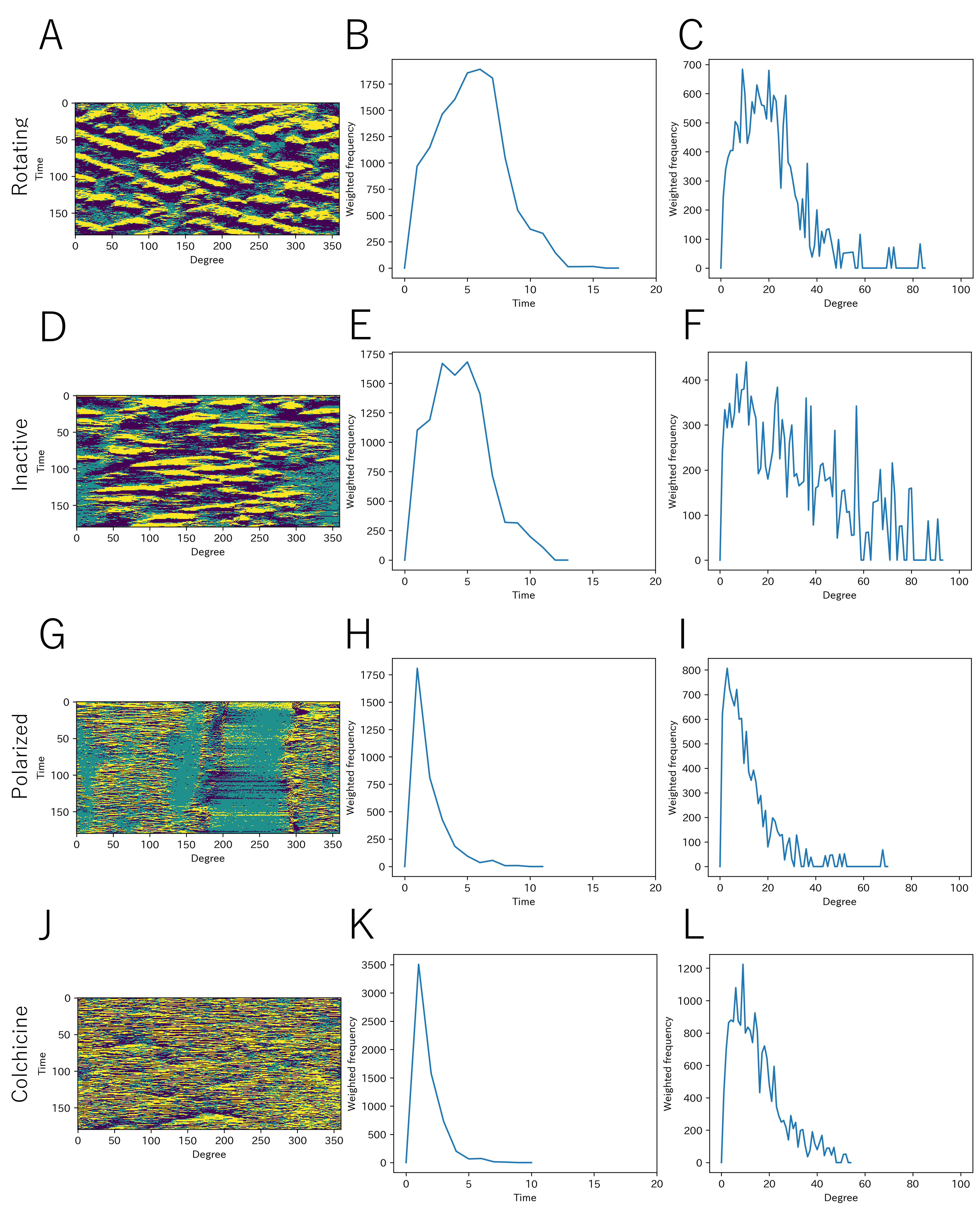}
    \caption{Protrusion dynamics of rotating, inactive, polarized, and colchicine-treated cell. (A,D,G,J). Yellow, black, and green represent protrusion, retraction, and stasis, respectively. (B,E,H,K) The weighted frequency of protrusion time.
(C,F,I,L) The weighted frequency of protrusion width (degree).}
    \label{fig:protrusion_maps}
\end{figure}

Next we made a kymograph of cell protrusion formation to quantify various characteristics (Fig. \ref{fig:protrusion_maps}). 
A two-dimensional array representing the relationship between angle and time was generated, where the states of protrusion, retraction, and stasis were assigned to each point. Protrusions are shown in yellow, retractions in black, and stasis in green.  

In the rotating cell, protrusive regions exhibit a consistent spatiotemporal characteristic length and propagate along the interface at a constant velocity (Fig. \ref{fig:protrusion_maps}A-C). The histogram of protrusion duration and the histogram of protrusion width are presented. When analyzing the frequency weighted by class, peaks were observed at approximately 5 frames for the temporal characteristic length and around 20\degree for the spatial characteristic length. These peaks can be interpreted as the duration and width of the protrusions, respectively.  

In the inactive cell, the characteristic length and number of protrusions, as well as rotational propagation, were slightly reduced (Fig. \ref{fig:protrusion_maps}D-F).  

In the polarized cell, retraction was prominently observed at approximately 180 \degree in the tail region, while protrusions in other areas appeared more fragmented (Fig. \ref{fig:protrusion_maps}G-I).  

In the colchicine-treated group, protrusions were present but finer in structure (Fig. \ref{fig:protrusion_maps}J-L).

% \begin{figure}
%     %\centering
%     \includegraphics[width=\linewidth]{_Figures/paper_spatiotemporal.png}
%     \caption{The control group. (A,D) time vs. degree array representing spatiotemporal proturusion dynamics,yellow for protrusion, black for retraction, and green for stasis . (B,E) The weighted frequency of protrusion time.
% (C,F) The weighted frequency of protrusion width(degree).}
%     %\label{fig:my_label}
% \end{figure}

% \begin{figure}
%     %\centering
%     \includegraphics[width=\linewidth]{_Figures/paper_colchicine_protrusion.png}
%     \caption{The colchicine-treated group. (A,D) time vs. degree array representing spatiotemporal proturusion dynamics,yellow for protrusion, black for retraction, and green for stasis . (B,E) The weighted frequency of protrusion time.
% (C,F) The weighted frequency of protrusion width(degree).}
%     %\label{fig:my_label}
% \end{figure}

\subsubsection{Measurement of surface tension}
% 曲率vs速度plotにおいて、s字の形をしていたものが、退縮部分のみ用いると直線に近い形になった。直線の傾きは平均15.6で、標準偏差は1.3であり、傾きの大きさが近い。 界面の法線速度V=σκに当てはめると、表面張力の係数σを抽出することができたと考えられる。
% ここでdx=0.30μm, dt=60secであり、σ=2.3*10^(-14)[m^2/s]である。
% colchicine処理群において突出の影響は抑制されており、曲率vs速度plotは直線かつ傾きの大きさが近い。直線の傾きは平均20.6で、標準偏差は3.8であった。このことから、傾きを表面張力の係数σとして捉えることができ、σ=3.1*10^(-14)[m^2/s]である。σはcontrol群から求めたものと多少異なるが、colchicineにより表面張力が影響されたことによるものと考えられる。

In the curvature vs. velocity plot, the initially S-shaped curve transformed into a nearly linear shape when only the retraction regions were considered. The slope of the resulting linear fit had a mean value of 15.6 with a standard deviation of 1.3, indicating consistency in slope magnitude. By applying the relationship \( V = \sigma \kappa \) for the normal interface velocity, the surface tension coefficient \( \sigma \) was estimated.  
Given that \( dx = 0.30 \, \mu\text{m} \) and \( dt = 60 \) sec, the estimated surface tension coefficient was \( \sigma = 2.3 \times 10^{-14} \, \text{m}^2/\text{s} \).  
In the colchicine-treated group, the influence of protrusions was suppressed, and the curvature vs. velocity plot exhibited a linear relationship with a consistent slope magnitude. The slope of the linear fit had a mean value of 20.6 with a standard deviation of 3.8. This suggests that the slope can be interpreted as the surface tension coefficient \( \sigma \), yielding \( \sigma = 3.1 \times 10^{-14} \, \text{m}^2/\text{s} \).  
Although the estimated \( \sigma \) differed slightly from that obtained in the control group, this variation is likely due to the effect of colchicine on surface tension.

% (A)曲率vs速度plotにおいて、s字の形をしているもの。(B)退縮部分のみから作成した曲率vs速度plot。(C)コルヒチン処理群における曲率vs速度plot。(C)コルヒチン処理群における曲率vs速度plot。
\begin{figure}
    %\centering
    \includegraphics[width=\linewidth]{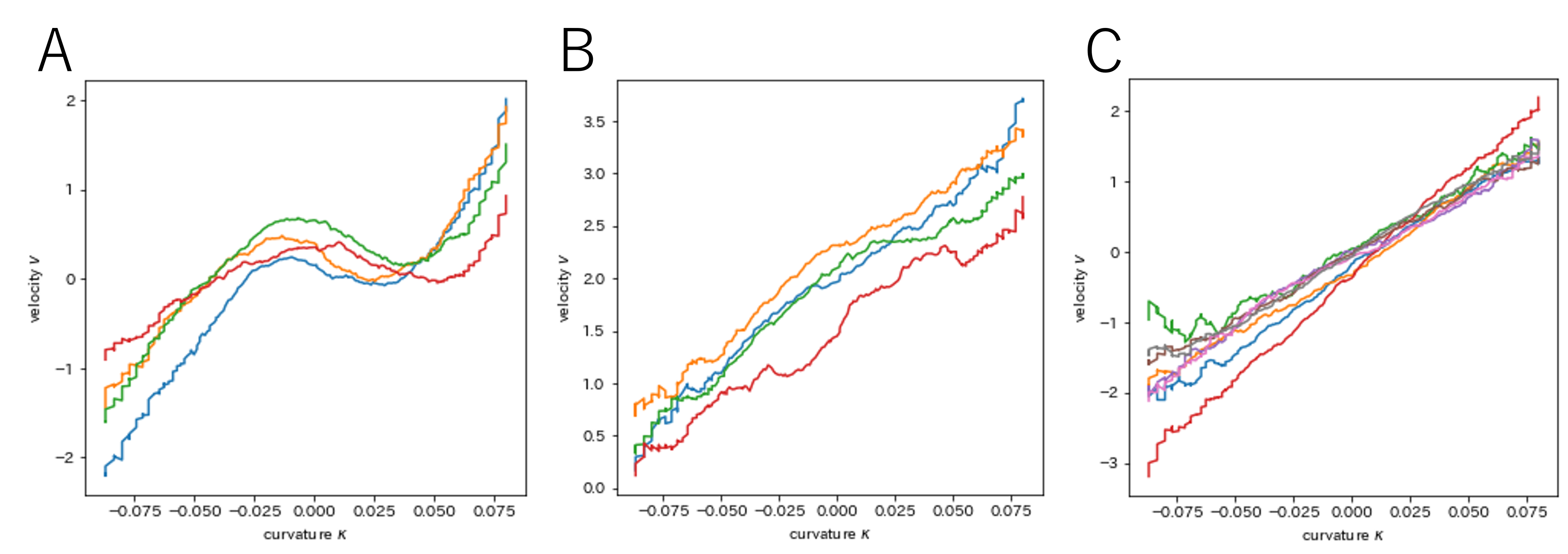}
    \caption{Estimation of surface tension using the curvature-velocity relationship of the cells. (A) Curvature vs. velocity plot exhibiting an S-shaped relationship. (B) Curvature vs. velocity plot generated using only the retraction regions. (C) Curvature vs. velocity plot for the colchicine-treated group.
}
\label{fig:experiment_surface_tension_measurement}
\end{figure}

% (A)曲率vs速度plotにおいて、s字の形をしているもの。(B)退縮部分のみから作成した曲率vs速度plot。(C)コルヒチン処理群における曲率vs速度plot。
% \begin{figure}
%     %\centering
%     \includegraphics[width=\linewidth]{_Figures/paper_experiment_s_shape_retraction_compare.png}
%     \caption{(A) Curvature vs. velocity plot exhibiting an S-shaped relationship.  (B) Curvature vs. velocity plot generated using only the retraction regions.
% }
%     %\label{fig:my_label}
% \end{figure}

% コルヒチン処理群における曲率vs速度plot。
% \begin{figure}
%     \centering
%     \includegraphics[width=0.8\linewidth]{_Figures/paper_colchicine_surface_tension.png}
%     \caption{Curvature vs. velocity plot for the colchicine-treated group.  
% }
%     %\label{fig:my_label}
% \end{figure}

\subsection{Mathematical model to reproduce single-cell movements.}
% 私たちは以下のprotrusionとsurface tensionの過程をn_substeps回繰り返すことで、細胞膜の突出とsurface tensionを実装するモデルを作成した。
% 細胞膜の表面張力をσ、突出幅をw_{p}、突出幅の増える速度をw^{+}_{p}、突出の回転速度をω_{p}、突出の成長速度をV_{p} 、突出の持続確率 をq_p、突出の生成確率をg_pとする。

We developed a model that implements cell membrane protrusions and surface tension by iterating the following protrusion and surface tension processes \( {n\_substeps} \) times.  

Let \( \sigma \) denote the surface tension of the cell membrane, \( w_{p} \) the protrusion width, \( w^{+}_{p} \) the rate of protrusion width increase, \( \omega_{p} \) the protrusion rotation velocity, \( V_{p} \) the protrusion growth velocity, \( q_p \) the protrusion persistence probability, and \( g_p \) the protrusion initiation probability.

\subsubsection{Protrusion Process}
% protrusion過程
% w_p= protrusion_num *dx, w^+_p =n_substeps* protrusion_plus_num *dx/dt, ω_p= n_substeps * rotate_num*dx/dt, V_p= n_substeps *dx/dt, q_p=persistence_p^ n_substeps, g_p= n_substeps * generate_pとする。

% 時刻tにおいて、outer boundaryから数点の突出の起点を指定する。
% persistence_pの確率で起点は時刻t-1から持続する。このとき、t-1の起点を中心としたカーネルとouter boundaryの交点の中から、真ん中の格子を選ぶことで、t-1の起点の近傍かつouter boundary上の格子を時刻tでの起点として指定する。起点をouter boundary上 でrotate_num個隣に指定することで、突出部分の回転を実装する。
% generate_pの確率で起点を、outer boundaryからランダムに指定する。この時、起点同士の距離が近くなりすぎないように、かつ起点での曲率が正でないように選ぶ。
% 2つの起点の距離が10未満になったとき、起点を2つとも消去する。
% 突出部がk回持続しているとき、起点を中心にprotrusion_num+(k-1)*protrusion_plus_num個の格子を1にすることで突出を実現する。

\begin{figure}
    \centering
    \includegraphics[width=\linewidth]{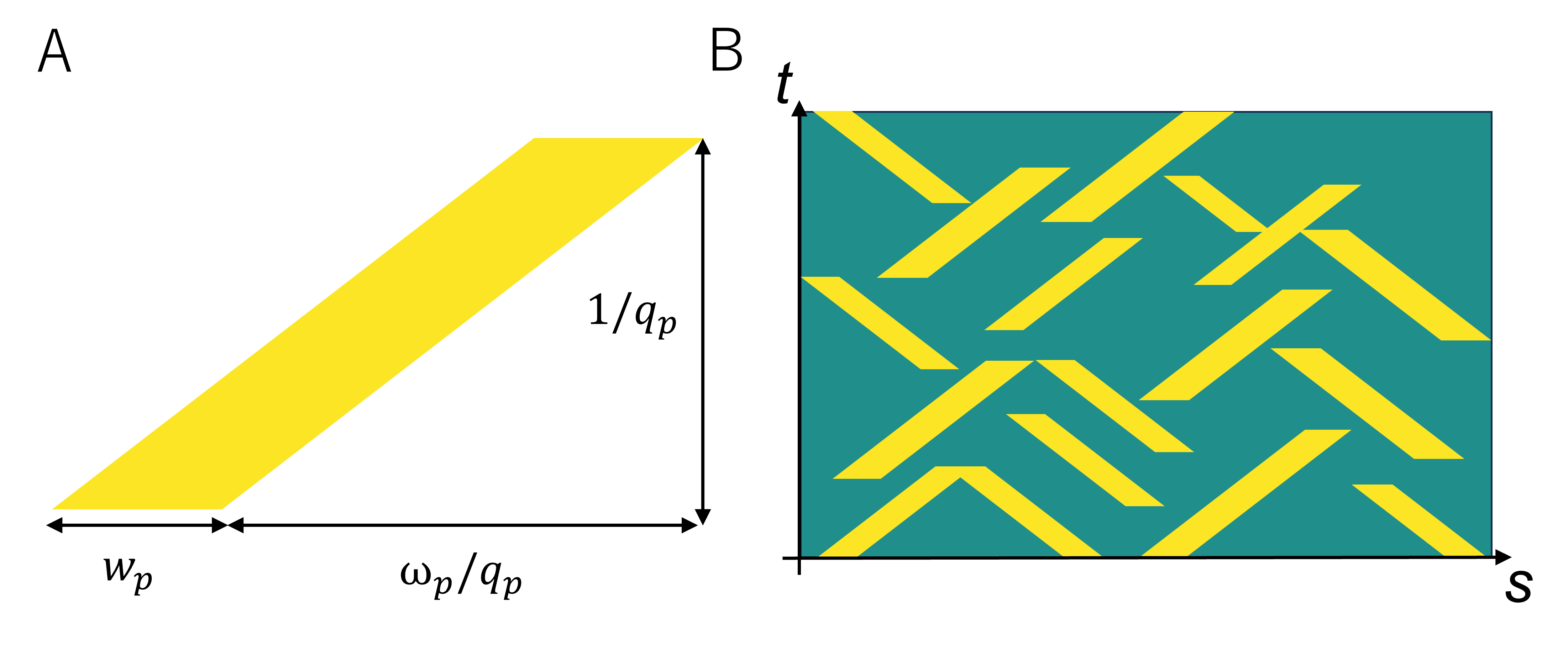}
    \caption{A diagram of the protrusion process. The horizontal axis represents the sequential index \( s \) assigned to the cell contour lattice points in a counterclockwise manner, while the vertical axis represents time \( t \). (A) Each parallelogram corresponds to the single protrusion formation, specified by lifespan ($1/q_p$), width ($w_p$), and speed ($\omega_p$). (B) The kymograph of the cell surface was implemented by multiple protrusion generation by probability $g_p$.}
    \label{fig:protrusion_scheme}
\end{figure}

We implement protrusion dynamics as follows (Fig. \ref{fig:protrusion_scheme}): at first, we define
\begin{align}
w_p &= \texttt{protrusion\_num} \times \texttt{dx},\\ 
w^+_p &= \texttt{n\_substeps} \times \texttt{protrusion\_plus\_num} \times \frac{\texttt{dx}}{\texttt{dt}},\\
\omega_p &= \texttt{n\_substeps} \times \texttt{rotate\_num} \times \frac{\texttt{dx}}{\texttt{dt}},\\
V_p &= \texttt{n\_substeps} \times \frac{\texttt{dx}}{\texttt{dt}}, \\
q_p &= \texttt{persistence\_p}^{\texttt{n\_substeps}}, \\
g_p &= \texttt{n\_substeps} \times \texttt{generate\_p}.
\end{align}
At time \( t \), several protrusion initiation points are selected from the outer boundary. With a probability of \( \texttt{persistence\_p} \), an initiation point persists from time \( t-1 \). In this case, among the intersections of the kernel centered at the initiation point from \( t-1 \) and the outer boundary, the central lattice point is chosen. This ensures that the initiation point at \( t \) remains near the previous location and is positioned on the outer boundary. The rotational movement of protrusions is implemented by shifting the initiation point by \( \texttt{rotate\_num} \) lattice points along the outer boundary. With a probability of \( \texttt{generate\_p} \), an initiation point is randomly selected from the outer boundary. To prevent excessive clustering, initiation points are chosen such that their distances do not become too small and that the curvature at the initiation point is not positive. If the distance between two initiation points is less than 10, both points are removed. If a protrusive region has persisted for \( k \) steps, the protrusion is implemented by setting \( \texttt{protrusion\_num} + (k-1) \times \texttt{protrusion\_plus\_num} \) lattice points to 1, centered at the initiation point.

% 突出過程の概念図を示す。横軸は、細胞輪郭の格子点を反時計回りに通し番号 s として割り当てたものを表し、縦軸は時間 t を表す。各平行四辺形は、突出領域が発生し、回転伝播して最終的に消失するまでのlifespanに対応する。

\subsubsection{Surface Tension Process}
% 界面張力の過程
% inner boundaryとouter boundaryの曲率を計算する。
% 曲率と細胞体積を用いて、ある格子において境界を移動させる確率pを計算する。
% V=σκ+α(ΔA/A_{0}), p=V /n_substeps * dt/dx
% ここで、σは表面張力、κは曲率、αは体積保存の強さ、A_{0}はtarget体積、ΔAは現在の体積とtarget体積との差を表す。
% 実際の計算では、無次元化したσとκから直接pを計算し、←いらない？消した
%αの値は体積の時間変化が極端でないように設定する。
% inner boundaryでp>0である格子は0にし、 outer boundaryでp<0である(負の方向に進む)格子は1にする。
% ある格子が斜め方向のみで他の格子と繋がっており、他の格子がさらに0に置き換えられて孤立した場合、その格子も0に置き換えて削除することとする。

\begin{figure}
    %\centering
    \includegraphics[width=\linewidth]{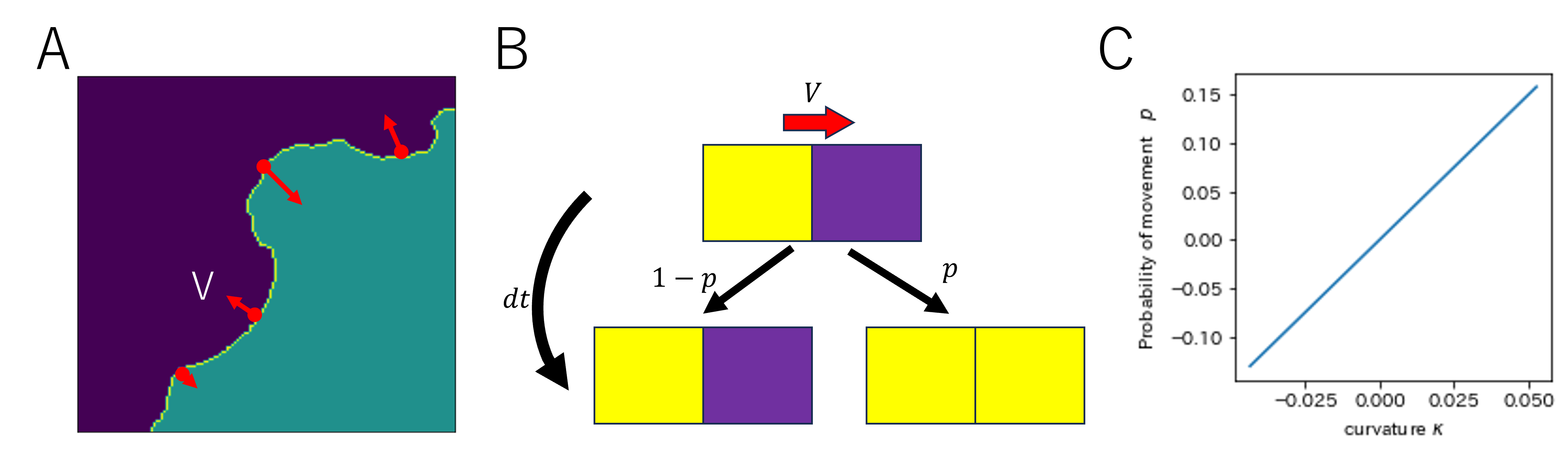}
    \caption{Implementation of surface tension process. (A) Movement of the interface due to surface tension (red arrows). (B) Implementation of interface movement. Non-cellular lattice points at the interface transition to cellular lattice points with a probability $p$. When the velocity of the interface is $V$, setting $p = V \, dt / dx$ ensures that the expected value of the interface velocity equals $V$.(C) Relationship between the curvature $\kappa$ at the interface and the probability $p$ of movement in the normal direction.}
    \label{fig:surface_tension_scheme}
\end{figure}

Next we implement surface tension as follows (Fig. \ref{fig:surface_tension_scheme}):
The curvature of both the inner and outer boundaries is computed. Using the curvature and cell volume, the probability \( p \) of moving a boundary lattice point is determined as follows:  
\[
V = \sigma \kappa + \alpha \left(\frac{\Delta A}{A_{0}}\right), \quad p = \frac{V}{\texttt{n\_substeps}} \times \frac{\texttt{dt}}{\texttt{dx}}
\]
where \( \sigma \) represents the surface tension, \( \kappa \) denotes the curvature, \( \alpha \) is the strength of volume conservation, \( A_{0} \) is the target volume, and \( \Delta A \) is the difference between the current volume and the target volume.  

The value of \( \alpha \) is chosen such that the temporal change in volume remains within a reasonable range.  

For the inner boundary, lattice points where \( p > 0 \) are set to 0, while for the outer boundary, lattice points where \( p < 0 \) (moving in the negative direction) are set to 1.  

If a lattice point is connected to other points only through diagonal connections, and if its neighboring points are further replaced with 0, the lattice point itself is also replaced with 0 to prevent isolation.

% (A) 表面張力による界面の移動（赤い矢印）。
% (B) 界面移動の実装。界面にある非細胞格子点が、確率 p で細胞格子点へと遷移する。界面の速度がVのとき、p=V dt/dxと設定することで、界面速度の期待値が V となるように制御される。
% (C) 界面における曲率κと法線方向速度Vの関係

\subsection{Quantitative measurement of our cell model}
\subsubsection{Curvature- velocity relationships for model-generated patterns.}
% 私たちのモデルで生成した細胞運動において、曲率vs界面速度のplotを示す。σ=5の界面張力のみを与えた場合、plotは傾きがほぼ5の直線となっている。これは与えた界面張力の強さを再現できている。界面張力とprotrusionを与えた場合、plotはprotrusionが相対的に強いほど、折れ線からS字カーブに変化した。

\begin{figure}
    %\centering
    \includegraphics[width=\linewidth]{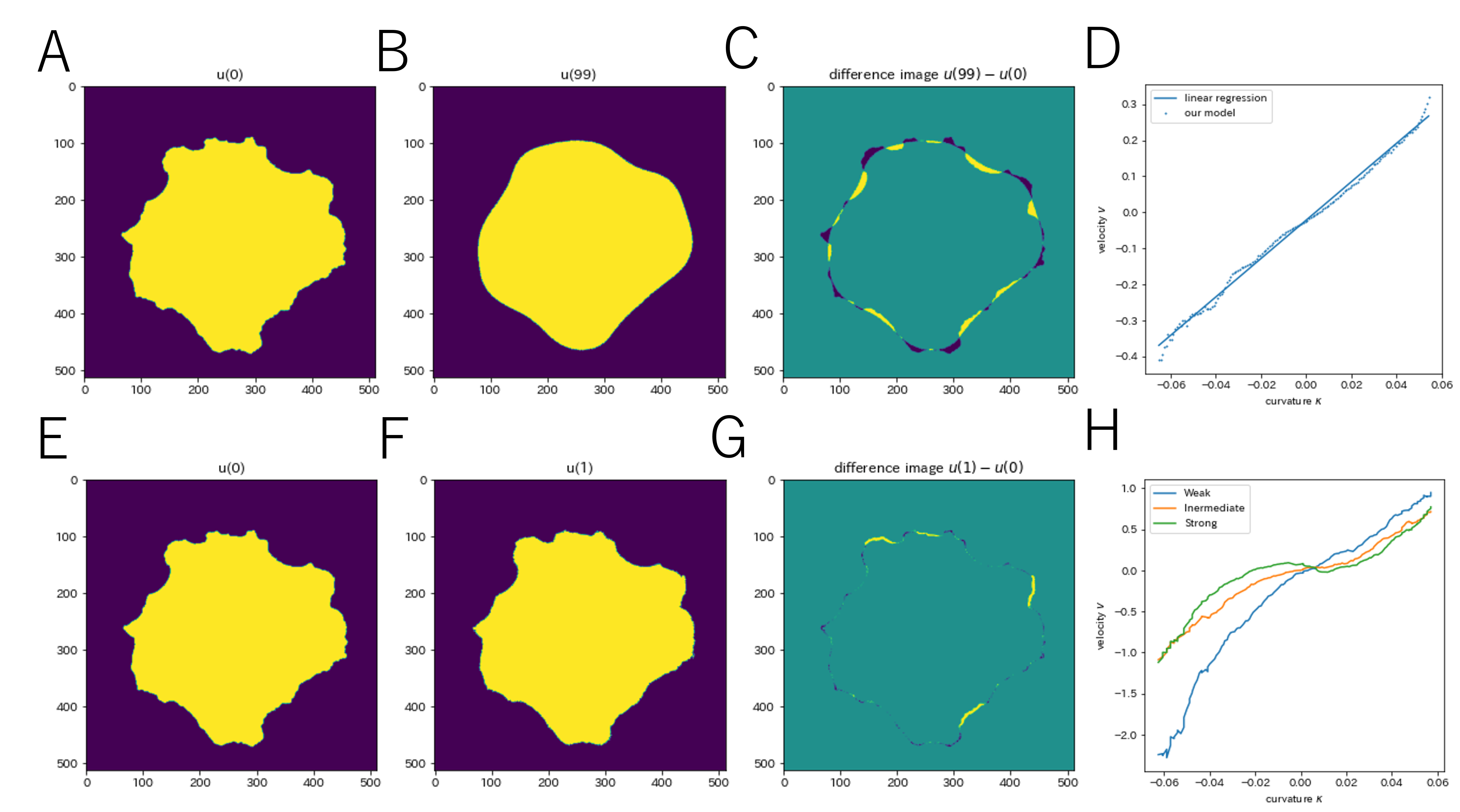}
    \caption{Cell dynamics and curvature-velocity relationship of our model. (A,D) Patterns of a single cell movement only by surface tension. \( u(t) \): Binarized cell image at time \( t \). (A) Cell image at \( t=0 \). (B) Cell image at \( t=99 \). (C) Difference image between \( t=99 \) and \( t=0 \). (D) Curvature vs. velocity plot of (A-B). The plot for \( \sigma = 5 \cdot dx^2/dt \) (blue dashed line) and its linear regression line (blue solid line) are shown. The model's setting of \( \sigma=5 \cdot dx^2/dt \) closely matched the measured slope (5.33). (E-H) Patterns of a single cell with surface tension and protrusion dynamics. \( u(t) \): Binarized cell image at time \( t \). (E) Cell image at \( t=0 \). (F) Cell image at \( t=1 \). (G) Difference image between \( t=1 \) and \( t=0 \). (H) Curvature vs. velocity plot obtained from (E,F). As protrusion activity increases, the plot transitions from a piecewise linear shape to an S-shaped curve. This result is consistent with the findings presented in Fig.\ref{fig:paper_singlecell_curvature_vs_velocity}A-D.}
 \label{fig:paper_model_images}
\end{figure}

We present the plot of curvature vs. interface velocity generated from the cell movement in our model (Fig. \ref{fig:paper_model_images}). When only surface tension with a strength of $\sigma=5 \cdot dx^2/dt$ was applied, the plot formed a nearly straight line with a slope of approximately 5, successfully reproducing the strength of the applied surface tension (Fig. \ref{fig:paper_model_images}A-D). When both surface tension and protrusion were applied, the plot transitioned from a piecewise shape to an S-shaped curve as the relative strength of protrusion increased (Fig. \ref{fig:paper_model_images}E-H).

\subsubsection{Measurement of surface tension}
% 曲率vs速度plotにおいて、s字の形をしていたものが、直線に近い形になった。また、直線の傾きは17.3で、 モデルにおいて設定した表面張力のパラメータσ=18*dx^2/dtと一致している。これにより、実験において測定した細胞でも、我々の手法で表面張力の影響を取り出すことができると考えられる。

\begin{figure}
    \centering
    \includegraphics[width=0.5\linewidth]{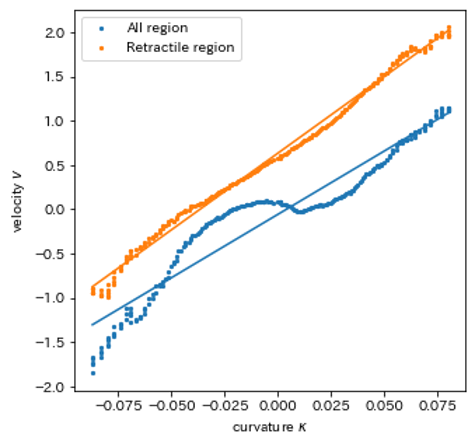}
    \caption{Curvature vs. velocity plots and their regression line. The blue line shows an S-shaped relationship, and the orange line shows a plot generated using only the retraction regions (orange).
}
    \label{fig:model_surface_measurement}
\end{figure}

In the curvature vs. velocity plot, the initially S-shaped curve transformed into a nearly linear shape (Fig. \ref{fig:model_surface_measurement}). The slope of the resulting linear fit was 17.3, which closely matched the surface tension parameter set in the model \( \sigma = 18 \cdot dx^2/dt \). This result suggests that our method can effectively extract the influence of surface tension even in experimentally measured cells.

% (A)曲率vs速度plotにおいて、s字の形をしているものと、その退縮部分のみを用いたもの。(B)退縮部分のみを用いたplotを線形回帰したときの傾き。(C)退縮部分のみを用いたplotを線形回帰したときのピアソンの積率相関係数。

\section{Discussion}
\subsection{Physiological importance of measuring surface tension}
% Mechanosensing(あまり書かれてない?)
% Cell division(書けそうではある？)

% 細胞膜の表面張力は、運動性、エンドサイトーシス、エキソサイトーシス、細胞形状の維持といったさまざまな細胞機能において重要な役割を果たします。膜張力が細胞の運動性と逆相関すると仮定して、細胞運動を説明する研究があります。また、膜張力がラメリポディアの突出を一方向に制限することで、細胞極性を維持するという研究があります。線虫精子細胞の膜張力を低下させることで、極性形成が妨げられ、運動速度が遅くなったという研究があります。

% また、膜張力が増加するとエキソサイトーシスが増やし、膜面積を供給するといった形で、張力の変化を緩衝することができます。これにより、細胞は機械的ストレスや環境変化に適応することができ、表面張力は細胞の柔軟性を支える重要な要素となっています。
% 膜張力を計測やそのメカニズムを調べることは、そうした細胞機能を理解する上で欠かせません。

Surface tension of the cell membrane plays a critical role in various cellular functions, including motility, endocytosis, exocytosis, and maintenance of cell shape \cite{Gauthier2012}. Some studies have proposed that membrane tension is inversely correlated with cell motility and used this relationship to explain cellular movement \cite{Ji2008, Keren2008}. Other research has suggested that membrane tension helps maintain cell polarity by restricting lamellipodial protrusion to a single direction. A study on nematode sperm cells has demonstrated that reducing membrane tension disrupts polarity formation and leads to a decrease in motility speed \cite{Batchelder2011}. In addition, increased membrane tension has been shown to enhance exocytosis, which supplies additional membrane surface area, thereby buffering changes in tension. This ability enables cells to adapt to mechanical stress and environmental changes, making membrane tension a critical factor in maintaining cellular flexibility \cite{Gauthier2011}.
Investigating the mechanisms and measurement of membrane tension is essential for understanding these cellular functions.

\subsection{Measurement method of membrane tension}

% 実験的には、光学的に捕捉されたビーズを用いて膜テザーを引っ張る方法で、膜張力を測定することができます。しかし、発生や集団的な細胞移動のように細胞へのアクセスが制限されているシステムでは測定が難しいです。タイムラプス画像によって膜張力を推定する方法は、細胞へのアクセスが制限されている状況でも可能です。また、他の実験と並行してタイムラプス画像を撮影することも可能であることから、簡便であるという利点もあります。

Experimentally, membrane tension can be measured by pulling membrane tethers using optically trapped beads \cite{Fazal2011}. However, this method is challenging to apply in systems where access to cells is restricted, such as during development or collective cell migration. Estimating membrane tension from time-lapse images offers a viable alternative in such scenarios. This approach is advantageous not only because it is feasible under conditions where direct access to cells is limited, but also because it can be performed alongside other experimental procedures, making it a convenient and versatile method.

\subsection{Advantages of the Proposed Model}
% 我々のモデルでは、界面張力を明示的に与えることができ、phase-field法と同じくらい曲率vs界面速度の関係において正確です。また、界面張力を0として設定することも可能で、計算が早いことも利点です。
% また、同じ離散的なモデルであるCPMと比べて、時間間隔や界面の速度といった物理量との対応がつきやすいという利点があります。物理量どうしを独立に設定することも可能です。

In our model, interfacial tension can be explicitly defined, achieving a level of accuracy comparable to the phase-field method in terms of the curvature vs. interface velocity relationship. Additionally, it is possible to set the interfacial tension to zero, and the model offers the advantage of faster computation.  
Compared to the Cellular Potts Model (CPM), which is also a discrete model, our approach has the advantage of providing a clearer correspondence between physical quantities such as time intervals and interface velocity. Furthermore, physical parameters can be independently specified.

\subsection{Limitation of our study}
% 我々のモデルは次のようなリミテーションを持つ。
% 格子の異方性が影響する可能性がある。これは空間刻みを増やすことで影響を減らすことができる。
% 変形に関する物理的なパラメータを直接設定しており、生化学反応のような内在する仕組みを示唆することができない。
% カーネル円よりも小さい構造は、正しく曲率を計測できない。この場合は、適切なカーネル円を取り直すか、空間座標から曲率を計算することが必要となる。

Our model has the following limitations: firstly, the influence of lattice anisotropy may affect the results. This effect can be mitigated by increasing the spatial resolution. 
Secondly, the model explicitly incorporates physical parameters related to deformation but does not account for intrinsic mechanisms such as biochemical reactions. 
Finally, structures smaller than the kernel circle cannot be accurately measured for curvature. In such cases, it is necessary to either redefine an appropriate kernel circle or calculate the curvature directly from spatial coordinates.

\section{Supporting information}
% \subsection{Curvature-velocity relationship in Allen-Cahn equation}
% % Allen-Cahn方程式で曲率に比例する速度で界面を動かし、この画像を使ってcurvature-velocity relationshipを算出すると、単純な直線ではなく階段状の曲線が見られる。この現象は以下の様に説明できる。離散性により、Allen-Cahn方程式$f(u))=u(1-u)(u-1/2)+d_u\delta u$における界面の速度は$V=d_u*1/r$である。$\delta x/V<dt$であれば界面の格子は動かない。曲率$\kappa, \kappa+\epsilon$の格子について、進む距離の差$d_u* \epsilon *dt<\delta x$であれば、2つの曲率の格子の速度は同じになる。このため、曲率vs界面速度relationshipにおいて、特定の曲率周辺で界面速度が一定になり、階段状の曲線となる。

% When the Allen-Cahn equation is used to move the interface at a velocity proportional to curvature, and the curvature-velocity relationship is calculated from the resulting images, a stepwise curve is observed instead of a simple straight line. This phenomenon can be explained as follows. Due to discretization, the velocity of the interface in the Allen-Cahn equation, 
% \[
% \frac{\partial u}{\partial t} = u(1-u)(u-1/2) + D_u \Delta u,
% \]
% is given by 
% \[
% V = D_u \cdot \kappa.
% \]
% If 
% \[
% V \cdot dt < dx,
% \]
% the interface grid point does not move. For lattice points with curvatures \( \kappa \) and \( \kappa + \epsilon \), the difference in the traveled distance, 
% \[
% D_u \cdot \epsilon \cdot dt,
% \]
% must satisfy 
% \[
% D_u \cdot \epsilon \cdot dt < dx.
% \]
% Under this condition, the grid points for these two curvatures will have the same velocity. Consequently, in the curvature-velocity relationship, the interface velocity remains constant around certain curvatures, resulting in a stepwise curve.

\subsection{Definition of phase field model}

% phase field法は、相を表す変数φ(r,t)についての連続場のモデルである。細胞を記述する場合は、細胞内をφ=1、細胞外をφ=0のように定義し、二相の間は薄い遷移層で連続に結ばれる。各相での自由エネルギーを表す二重井戸型ポテンシャルと界面エネルギー、体積保存、化学走化性などからなるエネルギーを導入し、エネルギーの汎関数微分としてφ(r,t)の時間発展を考える。反応拡散方程式と組み合わせることができ、多様な形状を表現することができるが、計算速度が遅いなどの問題がある。
% 非線形反応拡散方程式はパターン形成のモデルとして広く用いられてきた。その最も簡単な例として、単独の双安定反応拡散方程式であるAllen-Cahn方程式が知られている。ここで、0<m<1, ε<<1とする。解はΦ=0, Φ=1の２つの領域と、それをつなぐ遷移層をもつ。この遷移層の幅は有限だが小さく、ε→0のsharp interface limitによって界面として近似することで、界面の発展方程式を得られる。特にm=1/2であるとき、界面の法線速度Vは界面の平均曲率に比例し、これは平均曲率方程式と呼ばれて広く研究されている。Supporting informationで簡易的な導出に触れる。
The phase-field model is first described by \cite{Kobayashi:1993ww}.  For describing cells, the intracellular region is defined as \( \phi = 1 \), the extracellular region as \( \phi = 0 \), and the two phases are connected by a thin transition layer. The method incorporates energy terms such as a double-well potential representing the free energy in each phase, interfacial energy, volume conservation, and chemotaxis. The time evolution of \( \phi(\mathbf{r}, t) \) is determined by the functional derivative of these energies:
\[\tau 
\frac{\partial \phi(\mathbf{r}, t)}{\partial t} = -\frac{\delta G}{\delta \phi(\mathbf{r}, t)},
\]
where \( G \) is the energy functional. While the method can be combined with reaction-diffusion equations to express a variety of shapes, it has limitations in computational speed.

Nonlinear reaction-diffusion equations have been widely used as models for pattern formation. One of the simplest examples is the single bistable reaction-diffusion equation, known as the Allen-Cahn equation. Here, let \( 0 < m < 1 \) and \( \varepsilon \ll 1 \).

\begin{align}
\tau \frac{\partial \phi}{\partial t} = \varepsilon^2 \nabla^2 \phi + \phi (1 - \phi) \left( \phi - \frac{1}{2} + m \right)
\end{align}

The solution consists of two regions, \( \Phi = 0 \) and \( \Phi = 1 \), connected by a transition layer. While the width of this transition layer is finite, it is small, and by taking the sharp interface limit as \( \varepsilon \to 0 \), the transition layer can be approximated as an interface. This approximation leads to an interface evolution equation.

In particular, when \( m = 1/2 \), the normal velocity \( V \) of the interface is proportional to its mean curvature. This is known as the mean curvature equation, which has been extensively studied.

\subsection{Sharp interface limit of Allen-Cahn equation}

% 2次元のAllen-Cahn方程式を考える。ε<<1として、τ=bε, ｍ=εf/√2とおく。極座標系を使って、axi-symmetricを仮定すると、以下のようになる。界面の法線速度をV、平均曲率をκとおく。coordinate rをfactor εで引き伸ばし、速度Vのtraveling front solutionを仮定すると、次の式を得る。εについて、0次の項を比較する。これは静止フロント解をもつ。またΦ'は次のように変形できる。ΦとΦ'に代入して、1次の項を比較すると、Vを得る。極座標系の原点を曲率円の中心に取ることで、一般的な場合も同じ関係式を得る。
A simplified derivation of this relationship is provided by \cite{Kobayashi:1993ww}.
We consider the two-dimensional Allen-Cahn equation. 

\begin{align}
\tau \frac{\partial \phi}{\partial t} = \varepsilon^2 \nabla^2 \phi + \phi (1 - \phi) \left( \phi - \frac{1}{2} + m \right)
\end{align}

Let \( \varepsilon \ll 1 \), and define \( \tau = b\varepsilon \) and \( m = \varepsilon f / \sqrt{2} \). Assuming axial symmetry and using polar coordinates, the equation takes the following form. 

\begin{align}
b \varepsilon \frac{\partial \phi}{\partial t} = \varepsilon^2 \left( \frac{\partial^2 \phi}{\partial r^2} + \frac{1}{r} \frac{\partial \phi}{\partial r} \right) + \phi (1 - \phi) \left( \phi - \frac{1}{2} + \frac{\varepsilon f}{\sqrt{2}} \right)
\end{align}

Let \( V \) denote the normal velocity of the interface and \( \kappa \) the mean curvature. By stretching the radial coordinate \( r \) by a factor of \( \varepsilon \) and assuming a traveling front solution with velocity \( V \)

\begin{align}
\eta = \frac{1}{\varepsilon} \left( r - V t \right)    
\end{align}

we obtain the following equation

\begin{align}
    \frac{d^2 \phi}{d \eta^2} + \varepsilon \left( b V + \kappa \right) \frac{d \phi}{d \eta} + \phi (1 - \phi) \left( \phi - \frac{1}{2} + \frac{\varepsilon f}{\sqrt{2}}
    \right) = 0
\end{align}

By comparing the zeroth-order terms in \( \varepsilon \), we find that the equation admits a stationary front solution. 

\begin{align}
\phi'' + \phi (1 - \phi) \left( \phi - \frac{1}{2} \right) = 0 
\quad \rightarrow \quad 
\phi = \frac{1}{2} \left( 1 - \tanh \frac{\eta}{2\sqrt{2}} \right)
\end{align}

Additionally, \( \Phi' \) can be rewritten in the following form:

\begin{align}
\phi' &= -\frac{1}{2} \left( 1 - \tanh^2 \frac{\eta}{2\sqrt{2}} \right) \cdot \frac{1}{2\sqrt{2}} \\
&= -\frac{1}{\sqrt{2}} \cdot \frac{1}{2} \left( 1 - \tanh \frac{\eta}{2\sqrt{2}} \right) \cdot \frac{1}{2} \left( 1 + \tanh \frac{\eta}{2\sqrt{2}} \right) \\
&= -\frac{1}{\sqrt{2}} \phi (1 - \phi)
\end{align}

Substituting \( \Phi \) and \( \Phi' \) back into the equation and comparing the first-order terms in \( \varepsilon \), we derive the expression for \( V \):

\begin{align}
\varepsilon \{ \left( bV + \kappa \right) \phi' + \phi (1 - \phi) \frac{f}{\sqrt{2}}\} = 0
\end{align}
\begin{align}
\varepsilon \cdot \frac{1}{\sqrt{2}} \phi (1 - \phi) \{ - (bV + \kappa) + f \} = 0
\end{align}
\begin{align}
bV = f - \kappa
\end{align}

By placing the origin of the polar coordinate system at the center of the curvature circle, the same relationship can be derived for a general case \cite{Kobayashi:1993ww}.

\subsection{Relationship between lattice model and interface equation}
% 確率的に移動が決定される離散格子モデルにおいて、速度 \( c \) で移動する界面がどの程度の誤差を持つかを評価する。  

% 時間および空間解像度をそれぞれ \( n_t \) および \( n_x \) に分割した離散格子モデルを考え、速度 \( c \) を表現する。  
% このモデルにおける時間ステップは次のように与えられる：
% \[
% dt = \frac{1}{n_t},
% \]
% 空間ステップは：
% \[
% dx = \frac{1}{n_x}.
% \]

% このとき、移動確率 \( p \) は以下のように表される：
% \[
% p = \frac{c dt}{dx} = c \cdot \frac{n_x}{n_t}.
% \]

% n_tを大きく取れば、1単位時間内に \( k \) 格子点移動する確率はポアソン分布に従い：
% \[
% P(k; c \cdot n_x).
% \]

% \( c \cdot n_x \) がある程度大きければ、ポアソン分布は正規分布に近似される。また、\( n_x \) を大きくすると、標準偏差（SD）は次のように比例して減少する：
% \[
% \frac{1}{\sqrt{n_x}}.
% \]

We evaluate the extent of error that arises when representing an interface moving at velocity \( c \) in a discrete lattice model where movement is determined probabilistically.  

A discrete lattice model is considered, where the unit time and length are divided into \( n_t \) and \( n_x \) lattices, respectively, to represent the velocity \( c \).  
The time step $dt$ and spatial step $dx$  in this model are given by:
\begin{align}
dt = \frac{1}{n_t}, dx = \frac{1}{n_x}.
\end{align}

Under this formulation, the probability \( p \) of movement is expressed as:
\[
p = c \cdot \frac{dt}{dx} = c \cdot \frac{n_x}{n_t}.
\]

For sufficiently large \( n_t \), the probability of moving \( k \) lattice points within unit time follows a Poisson distribution:
\[
P(k; c \cdot n_x).
\]

When \( c \cdot n_x \) is sufficiently large, the Poisson distribution closely approximates a normal distribution. Furthermore, increasing \( n_x \) reduces the standard deviation (SD) proportionally as:
\[
\frac{1}{\sqrt{n_x}}.
\]

Therefore, the behavior of this system is similar to the interface equation if we take sufficiently large $n_x$ and $n_t$.

%\setcitestyle{maxnames=3}  % Limits author list in citations

\bibliographystyle{abbrvnat}
\bibliography{main.bib}

\end{document}